\documentclass[pra,reprint,superscriptaddress]{revtex4-2}
\usepackage[T1]{fontenc}
\usepackage{physics}
\usepackage{amsmath,amsfonts,bbm, graphicx, color,natbib}
\usepackage{amssymb}
\usepackage{appendix}
\usepackage{relsize}
\usepackage{xcolor}
\usepackage{scalerel}
\usepackage{tikz}
\usepackage{pdfpages}
\usetikzlibrary{svg.path}
\usepackage[normalem]{ulem}
\usepackage{amsfonts}
\usepackage{latexsym}
\usepackage{textcomp}
\usepackage[draft]{fixme}

\usepackage{listings}

\definecolor{orcidlogocol}{HTML}{A6CE39}
\tikzset{
	orcidlogo/.pic={
		\fill[orcidlogocol] svg{M256,128c0,70.7-57.3,128-128,128C57.3,256,0,198.7,0,128C0,57.3,57.3,0,128,0C198.7,0,256,57.3,256,128z};
		\fill[white] svg{M86.3,186.2H70.9V79.1h15.4v48.4V186.2z}
		svg{M108.9,79.1h41.6c39.6,0,57,28.3,57,53.6c0,27.5-21.5,53.6-56.8,53.6h-41.8V79.1z M124.3,172.4h24.5c34.9,0,42.9-26.5,42.9-39.7c0-21.5-13.7-39.7-43.7-39.7h-23.7V172.4z}
		svg{M88.7,56.8c0,5.5-4.5,10.1-10.1,10.1c-5.6,0-10.1-4.6-10.1-10.1c0-5.6,4.5-10.1,10.1-10.1C84.2,46.7,88.7,51.3,88.7,56.8z};
	}
}

\newcommand\orcid[1]{\href{https://orcid.org/#1}{\mbox{\scalerel*{
				\begin{tikzpicture}[yscale=-1,transform shape]
				\pic{orcidlogo};
				\end{tikzpicture}
			}{|}}}}
\usepackage[colorlinks]{hyperref}

\hypersetup{
	bookmarksnumbered,
	pdfstartview={FitH},
	citecolor={blue},
	linkcolor={blue},
	urlcolor={blue},
	pdfpagemode={UseOutlines}}
\definecolor{darkgreen}{RGB}{20,100,20}
\definecolor{darkblue}{RGB}{0,0,130}
\definecolor{darkred}{rgb}{.8,0,0}

\usepackage{soul}
\newcommand{\nn}{\nonumber}

\makeatletter
\AtBeginDocument{\let\LS@rot\@undefined}
\makeatother

\begin{document}
	
	\title{Effect of relativistic acceleration on tripartite entanglement in Gaussian states}
	
	\author{Jan A. Szypulski\orcid{0000-0003-1772-9618}}
	\email{j.szypulski@student.uw.edu.pl}
\affiliation{Institute of Theoretical Physics, University of Warsaw, Pasteura 5, 02-093 Warsaw, Poland}

	\author{Piotr~T.~Grochowski\orcid{0000-0002-9654-4824}}
	\email{piotr@cft.edu.pl}
	\affiliation{Center for Theoretical Physics, Polish Academy of Sciences, Aleja Lotnik\'ow 32/46, 02-668 Warsaw, Poland}
%	\affiliation{ICFO - Institut de Ci\`encies Fot\`oniques, The Barcelona Institute of Science and Technology, Av. Carl Friedrich Gauss 3, 08860 Castelldefels (Barcelona), Spain}
	
	\author{Kacper~D\k{e}bski\orcid{0000-0002-8865-9066}}
\email{kdebski@fuw.edu.pl}
\affiliation{Institute of Theoretical Physics, University of Warsaw, Pasteura 5, 02-093 Warsaw, Poland}
	
\author{Andrzej~Dragan\orcid{0000-0002-5254-710X}}
\email{dragan@fuw.edu.pl}
\affiliation{Institute of Theoretical Physics, University of Warsaw, Pasteura 5, 02-093 Warsaw, Poland}
\affiliation{Centre for Quantum Technologies, National University of Singapore, 3 Science Drive 2, 117543 Singapore, Singapore}
	\date{\today}
	
	\begin{abstract}
        Efficiencies of quantum information protocols can be affected by the noninertial motion of involved parties, e.g., in a gravitational field.
       Most of the previous studies focused on the effects of such a motion on a bipartite entanglement as a resource.
       We show how a tripartite Gaussian resource state of continuous variables is affected by the acceleration of the observers.
       Specifically, we analyze how genuine tripartite entanglement degrades due to the motion for various initial states and trajectories of the involved parties.
	\end{abstract}
	
	\maketitle
	
    	\section{Introduction}

All quantum information protocols require devices that can store and process quantum states.
Usually, the motion of these devices provides a negligible contribution to the overall efficiency of a given protocol, however, regimes in which relativistic effects can play a significant role are not far away from being achieved.
Specifically, all quantum states are observer-dependent~\cite{Takagi1986,Crispino2008}, a conclusion to which the community arrived after pioneering works involving the discoveries of Unruh and Hawking effects in 1970s~\cite{Fulling1973,Hawking1974,Hawking1975,Unruh1976,Unruh1977,Davies1975} and analysis of quantum information protocols in relativistic settings in the 1990s and 2000s~\cite{Peres2002,Alsing2002,Gingrich2002,Gingrich2003,Pachos2002,TERASHIMA2003,Peres2004,Czachor1997,Ahn2003,Massar2006,Ball2006}.

From the resource-oriented point of view, it was also shown that entanglement of the resource state is observer-dependent if the involved parties undergo a uniformly accelerated motion~\cite{Alsing2003b,Alsing2004,Fuentes2005,Alsing2006a,Bruschi2010,Bruschi2012a}.
A decrease of the quantum teleportation fidelity in such a setting compared to the inertial scenario strongly suggested that entanglement degradation goes beyond Lorentz mixing of degrees of freedom.
Apart from that, later studies confirmed that spatial degrees of freedom of global modes are entangled, also including the vacuum state~\cite{Reznik2003,Reznik2005,Lin2010,Olson2011}.
However, initial investigations addressed only global modes, missing the necessity of localization in space and time that is called for when any realistic protocol is to be performed.
Different approaches have been employed to tackle this problem---moving cavities~\cite{,Bruschi2012a,Friis2012,Friis2012a,Bruschi2013a,Friis2012b}, point-like detectors~\cite{Lin2008,Lin2008a,Lin2009,Lin2015,Doukas2010} and localized wave packets~\cite{Downes2013,Dragan2012,Dragan2013b,Doukas2013,Richter2017,Ahmadi2016,Grochowski2017,DD18,Grochowski2019}.

We focus on the last framework, established in~\cite{Ahmadi2016} and later generalized in~\cite{DD18,Grochowski2019} that provides a way to compute the effect of acceleration on multi-mode Gaussian states of localized wave packets in various geometries.
The entanglement-focused studies of moving quantum observers involved mostly investigations of two-party geometries, such as in the case of e.g. entanglement harvesting or effect of acceleration on efficiencies of two-party quantum protocols.
The interest in tripartite entanglement has been scarce, including extraction of genuine tripartite entanglement from the vacuum~\cite{Lorek2014} and purity loss for general multi-mode Gaussian states undergoing acceleration~\cite{DD18}.

The work aims to analyze how a tripartite Gaussian resource state of continuous variables is affected by the acceleration of the observers.
We explore symmetric channels in which each of the observers moves with the same, shared proper acceleration $\mathcal{A}$. 
As for the initial Gaussian states, fully symmetric and bisymmetric ones are considered, in each case showing relative degradation of the genuine tripartite entanglement (GTE), the stronger, the larger its initial value is.
We explicitly provide a numerical recipe for the calculation of GTE in these scenarios. 

The structure of the paper is as follows.
Section~\ref{framework} reintroduces the basics of the effect of acceleration on quantum states as a Gaussian channel and shortly characterizes three-mode Gaussian states.
In Section~\ref{sec:GTE} we provide measures of genuine entanglement for continuous variable (CV) tripartite states and overview the calculation algorithm.
Section~\ref{resu} contains the main results, involving analysis of acceleration on symmetric and bisymmetric states, while Section~\ref{conc} concludes the work.

\section{Framework}\label{framework}
\subsection{Gaussian channel}
Let us consider the real scalar massive quantum field $\hat{\Phi} $ in $1+1$ dimensional spacetime that satisfies the Klein-Gordon equation, $\left(\Box+m^2\right)\hat{\Phi}=0$, written in natural units of $c=\hbar=1$.
The field operator can be expressed in any coordinate system, but in our studies, we are focused on Minkowski coordinates corresponding to an inertial observer, and Rindler coordinates representing uniformly accelerated ones.
We restrict our considerations to the quantum states spanned by three orthonormal positive frequency solutions of the Klein-Gordon equation, i.e., we can decompose the field operator $\hat{\Phi}$ in two alternative ways:
\begin{equation}
\hat{\Phi}
=\sum_{i=1}^{3} \phi_i \hat{f}_i+\phi_i^* \hat{f}_i^{\dagger}=
\sum_{i=1}^{3} \psi_i \hat{d}_i+\psi_n^* \hat{d}_i^{\dagger},
\end{equation}
where $\phi_i$ is a solution of the field equation in the Minkowski frame, $\hat{f}_i$ and $\hat{f}_i^{\dagger}$ are corresponding annihilation and creation bosonic operators and $\psi_i$ is a solution in the Rindler frame and similarly, $\hat{d}_i$ and $\hat{d}_i^{\dagger}$ are corresponding annihilation and creation bosonic operators.

We will further restrict ourselves to the family of Gaussian quantum states, i.e., such ones that have Gaussian Wigner function and as such can be fully characterized by the first and second moments.
Moreover, It is worth noting that the transformation between two frames of reference of any quantum state can be described by a linear Bogolyubov transformation of creation and annihilation operators~\cite{Birrell1982}. However, in the case of Gaussian states, the whole operation of changing the reference frame can be seen as an action of a quantum Gaussian channel~\cite{Ahmadi2016} on the state prepared by the inertial observer.

Let us introduce a vector of quadrature operators, $\hat{\bm{X}}^{(\lambda)} \equiv\left(\hat{q}_1^{(\lambda)},\hat{p}_1^{(\lambda)},\hat{q}_2^{(\lambda)},\hat{p}_2^{(\lambda)},\hat{q}_3^{(\lambda)},\hat{p}_3^{(\lambda)}\right)^T $, where 
$\hat{q}_k^{(\lambda)}\equiv(\hat{\lambda}_{k}+\hat{\lambda}_{k}^{\dagger})/\sqrt{2}$ and 
$\hat{p}_k^{(\lambda)}\equiv(\hat{\lambda}_{k}-\hat{\lambda}_{k}^{\dagger})/\sqrt{2}i$.
Then, the first moment $\bm{X}^{(\lambda)}$ of a quantum state is given by an average value of the vector operator, $\bm{X}^{(\lambda)}=\langle\hat{\bm{X}}^{(\lambda)}\rangle$ \cite{gerardo3}.
The second moments form a covariance matrix (CM), \cite{gerardo3} $\sigma^{(\lambda)}_{kl}\equiv \langle \{ \hat{X}_k^{(\lambda)}-X_{k}^{(\lambda)},\hat{X}_l^{(\lambda)}-X_{l}^{(\lambda)} \} \rangle$, where $\{\cdot,\cdot\}$ is an anti-commutator. 
Here, $\lambda = f$ corresponds to the inertial modes, while $\lambda = d$ to the accelerated ones.

The action of a generic Gaussian channel is completely characterized by a pair of matrices $M$ and $N$ that act on the first and the second moment in the following way~\cite{Holevo2001}:
\begin{align}
\label{gausschannel}
\bm{X}^{(d)}&=M \bm{X}^{(f)}\mbox{,}
\nonumber
\\
\sigma^{(d)}&=M\sigma^{(f)}M^T+N\mbox{.} 
\end{align}
The exact form of the $M$ and $N$ matrices for a general multimode Gaussian state was derived in \cite{DD18}.
Here we are focused only on three-mode states so these matrices are both $6\times6$ dimensional.

\begin{figure}
\centering
\includegraphics[width=0.9\linewidth]{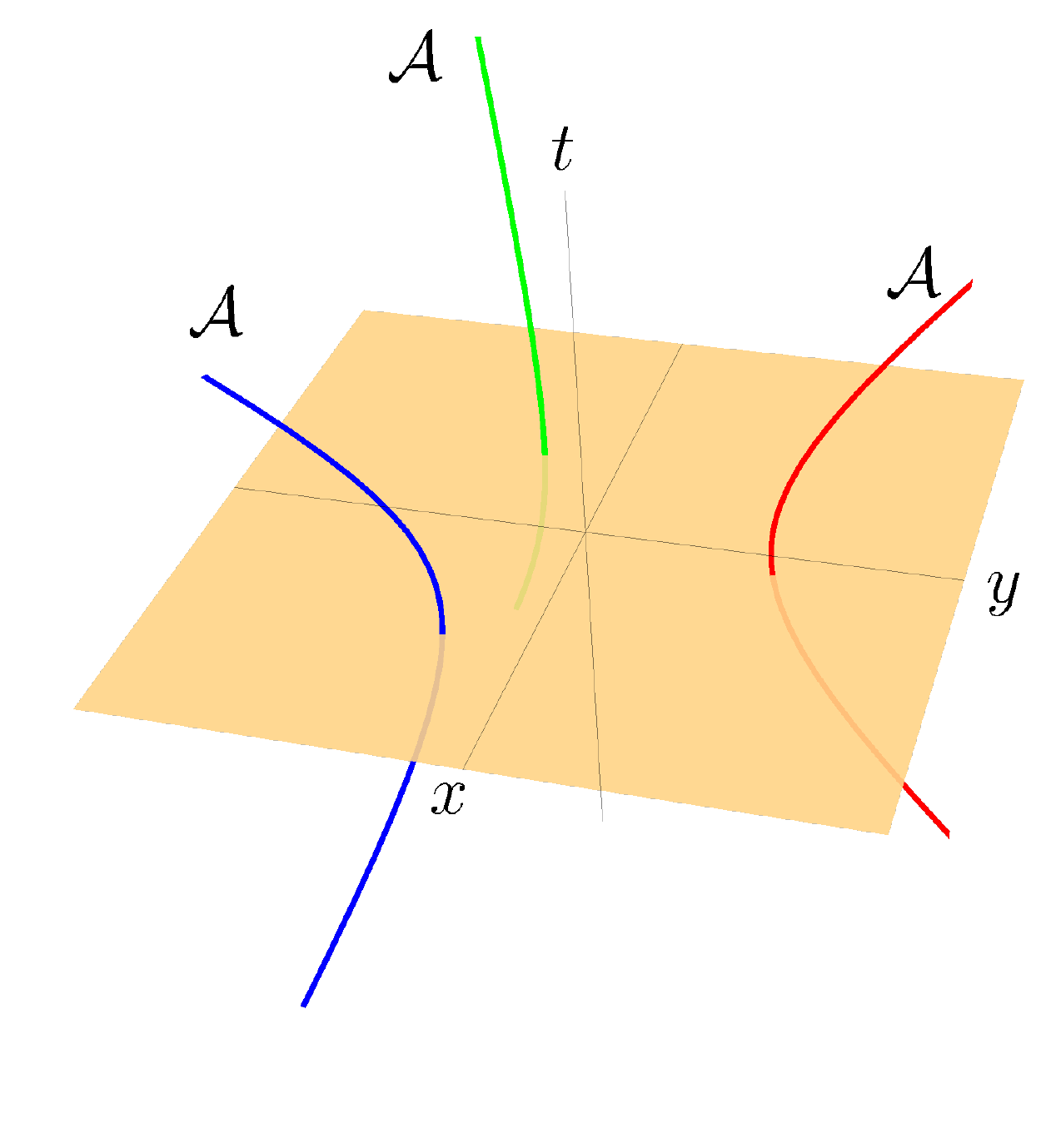}
\caption{Scheme of motion of three uniformly accelerating observes characterized by a shared proper acceleration $\mathcal{A}$.} 
\label{fig1}
\end{figure}

We now apply channel to the case of noninertial observers moving in three different directions (see Fig.~\ref{fig1}), each accelerating with identical proper acceleration $\mathcal{A}$.
Although the scheme discussed in \cite{DD18} involves only $1+1$ dimensional spacetime, a generalization to the higher dimensional case discussed in \cite{Grochowski2019} is straightforward.
The initial state spanned by three modes of the inertial observer is prepared in such a way that it can be also detected by three accelerating observers. Each of these accelerated observers will have access only to one mode of the field calculated in the Rindler reference frame. Together, these three observers measure the final state of the quantum field, and the difference between the final and the initial state of the quantum field can be interpreted as the result of the considered Gaussian channel.

\begin{figure}
\centering
\includegraphics[width=0.9\linewidth]{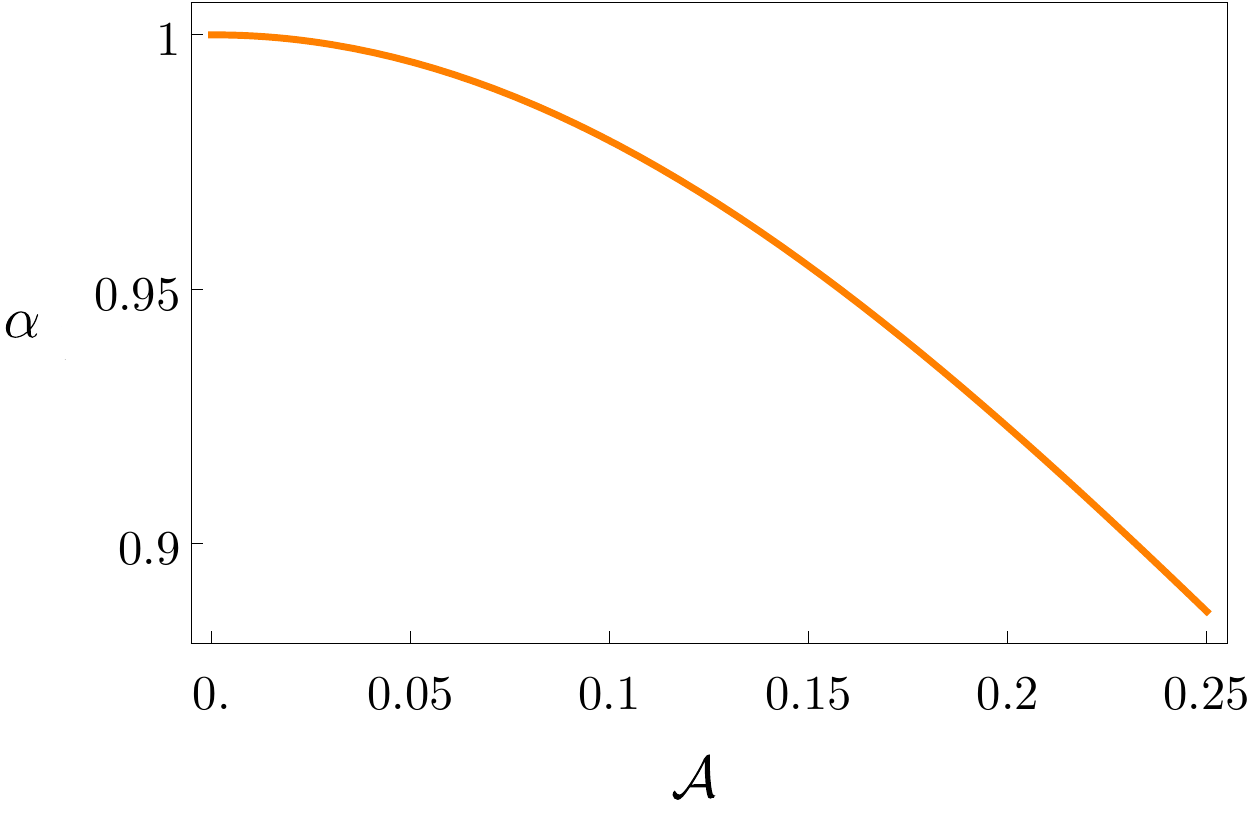}
\caption{The dependence of the overlap between inertial and accelerated modes as a function of the observer's proper acceleration.} 
\label{fig:przekrycie_od_a}
\end{figure}

For such an example of three observers, we can introduce the overlap coefficients,
\begin{align}
 \alpha_{ij}=& \delta_{i j}\left(\psi_i\middle|\phi_j\right)
=
\delta_{i j}\alpha
~~\mbox{,}
\nonumber
\\
\beta_{ij}=&-\delta_{i j}\left(\psi_i\middle|\phi_j^*\right)
=
\delta_{i j}\beta
~~\mbox{.}
\end{align}
Note that the overlap between inertial and accelerated mode is the same for all three observers.
It follows from the assumption of equal accelerations.
The particular values of the mode overlaps $\alpha$ and $\beta$ depend on the choice of mode functions and we stick to the ones used previously in~\cite{Ahmadi2016,Grochowski2017,DD18,Grochowski2019}.
For such a choice, $\beta$ coefficients are at least 8 orders of magnitude smaller than $\alpha$'s and can be safely neglected.
Then, the Gaussian channel is simplified to
\begin{align}
M &=\left( \alpha \mathbb{I}_{2\times2} \right)^{\bigoplus 3}
\nonumber
\\
N &=\left[\left( 1-\alpha^2\right) \mathbb{I}_{2\times 2} \right]^{\bigoplus 3}
\mbox{,}
\end{align}
where $\mathbb{I}_{2\times2}$ is an $2\times2$ identity matrix.
The numerical value of $\alpha$ coefficient is presented in Fig.~\ref{fig:przekrycie_od_a}. 

\subsection{Three-mode Gaussian states}
	For future convenience, let us define and write down the covariance matrix $\sigma_{123}$ of a three-mode Gaussian state in terms of two-by-two submatrices as
	\begin{equation} \label{eq:CM}
    \sigma_{123} = \begin{pmatrix}
        \sigma_1 & \epsilon_{12} & \epsilon_{13}\\
        \epsilon_{12}^T & \sigma_2 & \epsilon_{23} \\
        \epsilon_{13}^T & \epsilon_{23}^T& \sigma_3 
\end{pmatrix},
\end{equation}
    where $\sigma_i$ is the variance matrix of the $i$-th mode, and $\epsilon_{ij}$ is the covariance matrix between $i$-th and $j$-th mode.
    
    Let us recall that the form of the CM of any Gaussian state can be simplified through local (unitary) symplectic operations (that therefore do not affect the entanglement or mixedness properties of the state) belonging to $Sp^{\oplus n}_{2, \mathbb{R}}$. Such reductions of the CMs are called “standard forms”. For the sake of clarity, let us write the explicit standard form CM of a generic pure three-mode Gaussian state \cite{Adesso06}:
	
	\begin{equation} \label{eq:CMp}
    \sigma^p_{sf} = \begin{pmatrix}
        a_1 & 0 & \epsilon^+_{12} & 0 & \epsilon^+_{13} & 0 \\
        0 & a_1 & 0 & \epsilon^-_{12} & 0 & \epsilon^-_{13}\\
        \epsilon^+_{12} & 0 & a_2 & 0 & \epsilon^+_{23} & 0 \\
        0 & \epsilon^-_{12} & 0 & a_2 & 0 & \epsilon^-_{23} \\
        \epsilon^+_{13} & 0 & \epsilon^+_{23} & 0 & a_3 & 0 \\
        0 & \epsilon^-_{13} & 0 & \epsilon^-_{23} & 0 & a_3\\
\end{pmatrix},
\end{equation}
where
\begin{widetext}
\begin{equation}\label{def:epsilon}
\epsilon_{ij}^\pm =   \frac{\sqrt{\left[(a_i-a_j)^2-(a_k-1)^2\right] \left[(a_i-{a_j})^2-({a_k}+1)^2\right]}\pm\sqrt{\left[({a_i}+{a_j})^2-({a_k}-1)^2\right] \left[({a_i}+{a_j})^2-({a_k}+1)^2 \right]}}{4 \sqrt{a_i a_j}}.
\end{equation}
\end{widetext}
With this parametrization, it becomes clear that CMs of pure states are fully determined by local single-mode symplectic eigenvalues $a_l$ (later we will sometimes refer to them as local mixednesses $a_l$), and going further, they also determine the genuine tripartite entanglement (cf. Section \ref{sec:GTE}). 
Defining the parameters
\begin{equation}
    a'_l \equiv a_{l} - 1,
\end{equation}
the Heisenberg uncertainty principle for single-mode states reduces to 
\begin{equation}
    a'_l\geq 0\quad \forall l=1,2,3   
\end{equation}
and the range of the allowed values local becomes a triangle inequality
\begin{equation}
    | a'_i - a'_j | \leq a'_k \leq a'_i + a'_j \label{eq:triangle}
\end{equation}
The full description of these facts can be found in \cite{Adesso06}. 

	\section{Tripartite entanglement in Gaussian states} \label{sec:GTE}
The key to understanding tripartite entanglement is the sharing inequality (sometimes called monogamy inequality or CKW inequality named after Coffman, Kundu, and Wooters~\cite{CKW00}).
It was introduced for systems of three qubits and later extended to systems of $n$ qubits by Osborne and Verstraete~\cite{OV06}.
The CKW sharing inequality for a three-party system looks as follows
	\begin{equation}
	    E^{i|(jk)} - E^{i|j} - E^{i|k} \geq 0, \label{monIneq}
	\end{equation}
	where $i, j, k$ numerate the three elementary particles and $E$ refers to a proper measure of bipartite entanglement. 
	
	It is natural to expect that inequality~(\ref{monIneq}) should hold for the states of CV systems as well, despite the fact that they are defined on infinite-dimensional Hilbert spaces and can in principle achieve infinite entanglement. 
	Therefore, the crucial ingredient to construct the CV version of CKW monogamy inequality is a proper measure of entanglement $E$, able to capture the difference between couple-wise and tripartite correlations.
	For qubit systems, there are different proposals.
	The tangle \cite{OV06} was used as a starting point for G. Adesso and F. Illuminati~\cite{GAtangle}, who gave the CV analog and proved the monogamy inequality (\ref{monIneq}) for all Gaussian states of three modes and all symmetric Gaussian states with an arbitrary number of modes. Hence, using the approach from~\cite{GAtangle}, we recall the notation and steps leading to the definition of the continuous variable tangle, which will provide us with a measure of GTE in CV systems. 
	
	\subsection{The continuous-variable tangle}
	    Given a generic pure state $\ket{\psi}$ of a $(1+N)$-mode CV system, we can write CV tangle $E_\tau$ as follows \cite{GAtangle}
	    \begin{equation}
	        E_\tau(\ket{\psi}) \equiv \ln^2 \Vert \tilde{\rho}\Vert_1, \quad \rho=\ket{\psi}\bra{\psi}, \label{def:tangle}
	    \end{equation}
	    where $\tilde{~}$ denotes partial transpose of the state and $\vert \cdot \vert_1$ is trace norm \cite{EisertPhD}. 
	    This is a proper measure of bipartite entanglement, being a convex, increasing function of the logarithmic negativity $E_\mathcal{N}$, which is equivalent to the entropy of entanglement for arbitrary pure states.
	    For a pure Gaussian state $\ket{\psi}$ with CM $\sigma^p$, it was found~\cite{GAtangle}, that
	    \begin{equation}
	        E_\tau (\sigma^p) = \text{arcsinh}^2\left(\frac{\sqrt{1-\mu_1^2}}{\mu_1}\right), \label{tangle}
	    \end{equation}
	    where $\mu_1=1/\sqrt{\det\sigma_1}$ (see (\ref{eq:CM})) is the local purity of the reduced state of the mode 1, described by a CM $\sigma_1$. 
	    Note that we consider $1\times n$ bipartion of the entanglement.
	    The restrictions of only pure states for the measure (\ref{def:tangle}) can be be smoothly extended to generic mixed states, with density matrix $\rho$, through the convex-roof formalism \cite{OS04}.
	    Consequently,  
	    \begin{equation}
	        E_\tau(\rho)\equiv \inf_{\{p_i, \psi_i\}}\sum_i p_i E_\tau(\psi_i), 
	    \end{equation}
	    where the infimum is taken over all convex decompositions of $\rho$ in terms of pure states $\{\psi_i\}$ with probability $p_i$ of measuring them.
	    If the index $i$ is continuous, the sum is replaced by an integral, and the probabilities $\{p_i\}$ by a probability distribution $\pi(\psi)$. 
	    
	    Any multi-mode mixed Gaussian state with CM $\sigma$ admits a decomposition in terms of pure Gaussian states only.
	    The infimum of the average contangle, taken over all decompositions into pure Gaussian states with a CM $\sigma^p$, defines the Gaussian contangle $G_\tau$ \cite{Adesso06}
	    \begin{equation}
	        G_\tau(\sigma) \equiv \inf_{\{\pi(d\sigma^p),~\sigma^p\}} \int \pi(d\sigma^p)E_\tau(\sigma^p).
	    \end{equation}
	The convex-roof construction brings one more important notion.
	The Gaussian contangle $G_\tau(\sigma)$ is an upper bound to the true contangle $E_\tau(\sigma)$, because $E_\tau$ can be in principle minimized over a non-Gaussian decomposition, i.e., 
	\begin{equation}
	    E_\tau(\sigma) \leq G_\tau(\sigma). 
	\end{equation}
	It can be shown that $G_\tau(\sigma)$ is a bipartite entanglement monotone under Gaussian local operations and classical communication (GLOCC) (for proof see \cite{AI05, WG04}).
	Finally, having all the preliminaries in mind, the Gaussian contangle can be expressed in terms of CMs as
	   \begin{equation}
	       G_\tau(\sigma ) = \inf_{\sigma^p\leq \sigma} E_\tau(\sigma^p), \label{Ginf}
	   \end{equation}
    where the infimum runs over all pure Gaussian states with condition $\sigma^p\leq \sigma$, which means that the matrix $\sigma-\sigma^p$ has to be semi-positively defined.
    For all pure Gaussian states of $1\times n$ bipartitions, $E_\tau=G_\tau$ holds.

	\subsection{Genuine tripartite entanglement via residual contangle }
	
\begin{figure}[h]
	    \centering
	    \includegraphics[width=0.9\linewidth]{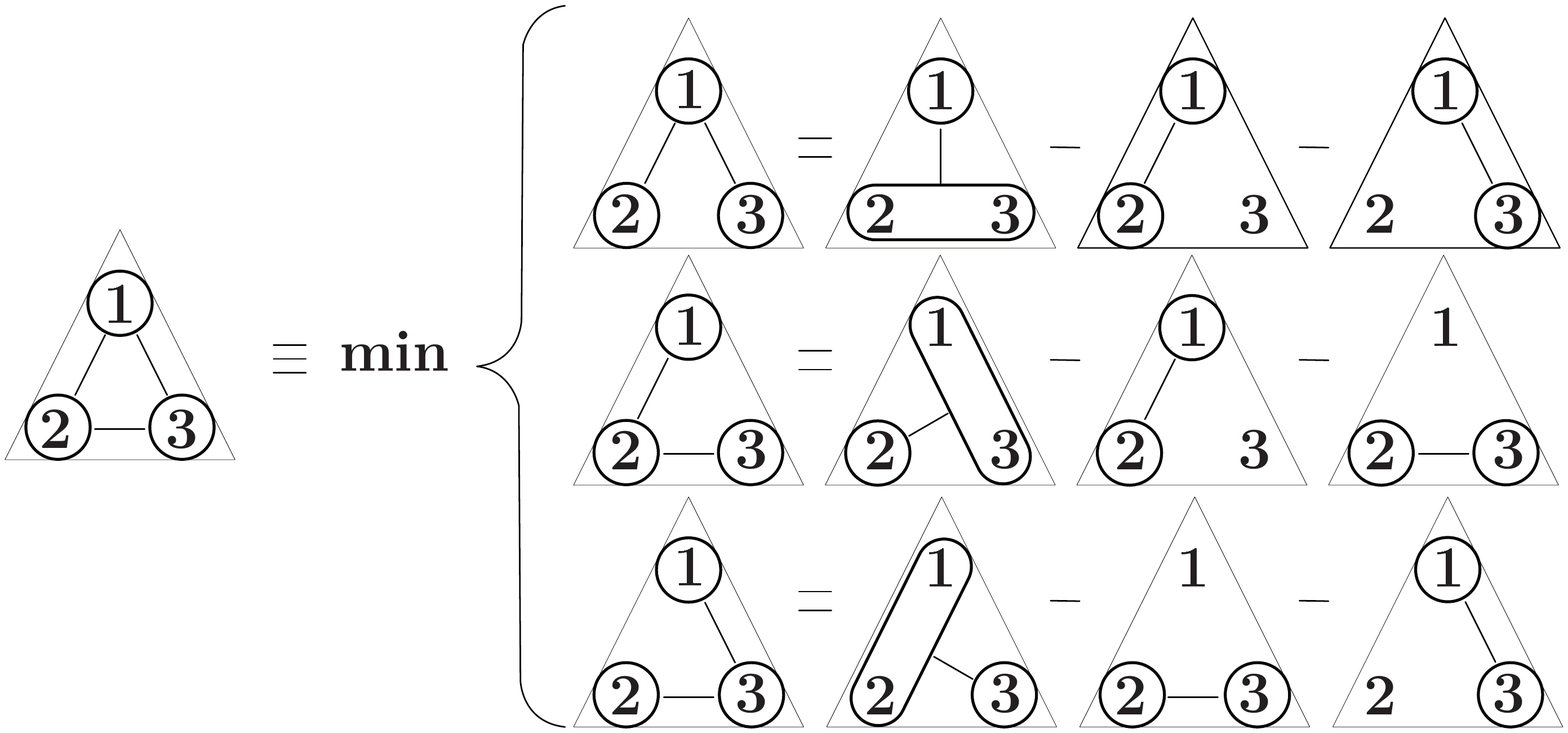}
	    \caption{Pictorial representation of minimal residual contangle.}
	    \label{fig:residaulContangle}
	\end{figure}

	Sharing constraint (\ref{monIneq}) provides a natural quantifier of genuine tripartite entanglement in systems of three qubits and much the same way in three-mode Gaussian states.
	As in the qubit case, the residual contangle is partition-dependent.
	It can take different values according to the choice of the reference mode, with the obvious exception of the fully symmetric states.
	The real quantification of tripartite entanglement is provided by the minimum residual contangle \cite{GAtangle}
	\begin{equation}
	    E_\tau^{i|j|k}\equiv\min_{(i,j,k)}\left[E_\tau^{i|(jk)} - E_\tau^{i|j}-E_\tau^{i|k}\right], \label{eq:residualContangle}
	\end{equation}
	where $(i,j,k)$ denotes all permutations of the three mode indices.
	This way of definition ensures that $E_\tau^{i|j|k}$ stays invariant under all permutations of the modes and is thus a genuine three-way property of any three-mode Gaussian state.
	See Fig. \ref{fig:residaulContangle} for pictorial representation. 
	One can adopt an analogous definition to (\ref{eq:residualContangle}) and obtain the minimum residual Gaussian contangle $G^{res}_\tau$,
	\begin{equation}
	    G_\tau^{res}\equiv G_\tau^{i|j|k}\equiv\min_{(i,j,k)}\left[G_\tau^{i|(jk)} - G_\tau^{i|j}-G_\tau^{i|k}\right], \label{def:Gres}
	\end{equation}
	which we will recognize as the genuine tripartite entanglement measure for Gaussian states. 
	
    For mixed Gaussian states, the definition of $(\ref{Ginf})$ can be extended over Gaussian contangle, and thus we have
    \begin{equation}
        G^{res}_\tau(\sigma)=\inf_{\sigma \geq \sigma^p}G^{res}_\tau(\sigma^p). \label{Gmixed}
    \end{equation}
    This constitutes a complex, nontrivial optimization problem.
    Later we show that for certain families of Gaussian states it is possible to find an efficient way of this optimization.
	
	\subsection{Calculation of genuine tripartite entanglement }
	
	We will now present the complete procedure to calculate the genuine tripartite entanglement in a pure three-mode Gaussian state $\sigma^p$ \cite{GAtangle}. 
	\begin{enumerate}
    \item {\bf Determine the local purities.} The state is globally pure ($\det \sigma^p = 1$), therefore, the only quantities needed for the computation of the tripartite entanglement are the three local mixednesses $a_l$ of the single-mode reduced states $\sigma_l$, $l = 1, 2, 3$ (see Eq. (\ref{eq:CMp})). 
    
    \item {\bf Find the minimum.} The minimum in the definition (\ref{def:Gres}) of the residual Gaussian contangle $G^{res}$ is attained in the partition where the bipartite entanglements are decomposed choosing as reference the mode $l$, which is the smallest among local mixednesses, $a_l \equiv a_{min}$.
    
    \item {\bf Check the range and compute.} Given the mode with the smallest local mixedness $a_{min}$ (say, for instance, mode 1) we can calculate the genuine tripartie entanglement. If $a_{min} = 1$, then mode 1 is uncorrelated from the others and $G^{res} = 0$. If, instead, $a > 1$ then we can get fully analytical expression for $G^{res}_{\tau}(\sigma^p)$ in Eq. (\ref{eq:GTE}), for details see Appendix \ref{CalculationOfGTE}. 
    \end{enumerate}

    For mixed states, the calculation of GTE can not be performed analytically for any random state. 
    Cumbersome expressions for the Gaussian contangles between modes $1|2,$ $1|3$ can be written, but the optimization of the $1|(23)$ bipartite Gaussian contangle has to be solved numerically \cite{AI05}. 
    However, in this paper, we show that for some families of Gaussian states there is an efficient way for the quantification of GTE.

\section{Effect of acceleration} \label{resu}
    
 In this section, we will show how the acceleration causes the decay of entanglement in three-mode, fully symmetric, or bisymmetric Gaussian states. 
    
    \subsection{Fully symmetric states}
    A fully symmetric three-mode pure squeezed vacuum state can be written as \cite{Adesso08,Serafini05}:
\begin{equation} \label{eq:CMvacuum}
    \sigma^p_{s} = \begin{pmatrix}
        \beta & \zeta &  \zeta \\
        \zeta & \beta & \zeta  \\
        \zeta &  \zeta & \beta 
\end{pmatrix},
\end{equation}
with
\begin{equation}
    \beta = \begin{pmatrix} b & 0 \\
    0 & b\end{pmatrix}, \quad \zeta= \begin{pmatrix} z_1 & 0 \\
    0 & z_2\end{pmatrix}, \label{betandzeta}
\end{equation}

\begin{equation}
    b=\frac{1}{3}\sqrt{4\cosh(4r)+5} \label{def:b},
\end{equation}
\begin{equation}
    z_1=\frac{2\sinh^2(2r)+3\sinh(4r)}{3\sqrt{4\cosh(4r)+5}}, \label{def:z1}
\end{equation}
\begin{equation}
    z_2=\frac{2\sinh^2(2r)-3\sinh(4r)}{3\sqrt{4\cosh(4r)+5}}\label{def:z2}.
\end{equation}

    One may prove that $(\ref{eq:CMp})$ and $(\ref{eq:CMvacuum})$ describe the same family of fully symmetric states, using two equivalent parametrizations.
    For details, see Appendix \ref{TwoParametrizations}.

 It is worth noting that fully symmetric pure states describe not only squeezed vacuum but also provide CV extensions of GHZ and W states. 
    Both these states are fully symmetric, i.e., changing the places of any pair among the three qubits will not change the quantum state. 
    To define an analog CV-state to $\ket{\psi_{GHZ}}$ \cite{GHZ90},
        \begin{equation}
        \ket{\psi_{GHZ}}=\frac{1}{\sqrt{2}}\bigg(\ket{000}+\ket{111}\bigg),
    \end{equation}
    one aims to maximize genuine tripartite entanglement. 
    On the other hand, to obtain an analog of $\ket{\psi_W}$ state~\cite{Dur2000ThreeWays},
        \begin{equation}
        \ket{\psi_W}=\frac{1}{\sqrt{3}}\bigg(\ket{100}+\ket{010}+\ket{001}\bigg)
    \end{equation}
     in CV-system one rather aims at maximizing the bipartite entanglement between any pair of three qubits \cite{Adesso06, Dur2000ThreeWays}. 
     
 Since pure Gaussian states are fully characterized by their local mixdnesses $\{a_l\}$, the only way to preserve the full symmetry, such that any permutation of state modes results in the same state, is to set all mixednesses $a_l$ equal to each other. 
    
    \subsection{Simplified numerical procedure}
         
    The family of fully symmetric states has one more surprising property. 
    All CMs of symmetric states, parametrized by vacuum squeezing $r$ commute, i.e., \begin{equation}
        [\sigma^p_s(r_1), \sigma^p_s(r_2)]=0, \quad \forall r_1, r_2 \in \mathbb{R}_+, 
    \end{equation}
    which means that they have a common eigenbasis.
    We use this common eigenbasis to simplify the computation of the effect of acceleration on entanglement. 
    The eigenvalues of any matrix $\sigma^{p}_s$ can be written in a simple form:
    \begin{eqnarray}
        &&\lambda_1=(b-z_1), \quad \lambda_3=(b+2z_1), \\ 
        &&\lambda_2=(b-z_2), \quad \lambda_4=(b+2z_2).
    \end{eqnarray}
    where $\lambda_1$ and $\lambda_3$ are repeated (double eigenvalues).
    For exact expressions, see Appendix \ref{LambaSym}.

    The Gaussian channel (\ref{gausschannel}) transforms the CM linearly and as a consequence its eigenvalues become 
    \begin{equation}
        \lambda^{(d)}_i=\alpha^2\lambda^{(f)}_i+(1-\alpha^2) \label{lambdaTransformed}
    \end{equation}
    where $\lambda_i^{(f)}$ is the $i$-th eigenvalue of an initial state and $\lambda^{(d)}_i$ is the eigenvalue of a state seen by accelerating observer.
    The superscripts $(d), (f)$ refer to $\hat{d}, \hat{f}$ operators.
    We know that due to the transformation the state losses its purity, thus in order to compute GTE of a state $\sigma$ seen by Rindler observer we need to use formula ($\ref{Gmixed}$).
    Then, the minimization over set of pure states with CM $\sigma^p$ needs to fulfill the relation that difference of matrices $\sigma-\sigma^p$ stays a semi-positively defined matrix,
    \begin{equation}
        \sigma^{(d)} - \sigma^{(d),p}\geq0. \label{eq:positivity}
    \end{equation}
    Here, we assume that symmetric Gaussian channel (\ref{gausschannel}) does not disturb the symmetry of the state.
    Thus, we narrow down the matrix search to symmetric state matrices. 
    In the common eigenbasis, this reduces to four inequalities of the following type: 
    \begin{equation}
        \lambda_i^{(d)}-\lambda_i^{(d),p}\geq0. \label{semicondition}
    \end{equation}
    For the initial state, the GTE is monotonically growing with squeezing $r$.
    Therefore, to find the infimum of ($\ref{Gmixed}$), we look for the smallest $r$ for which all inequalities (\ref{semicondition}) are satisfied. 
    It turns out that we can even narrow our computation to consideration of one eigenvalue.
    Some minimal $r_m$ saturates the inequalities (\ref{semicondition}), which gives
    \begin{equation}
        \alpha^2(\lambda_1(r)-1)-(\lambda_1(r_{m})-1)=0 \label{eq:lambda1solve}
    \end{equation}
    and leaves the rest of inequalities unsaturated
    \begin{equation}
        \lambda_j^{(d)}-\lambda_j^{(d),p}>0, \quad j\neq 1. 
    \end{equation}
    We have not managed to find appropriate analytic proof for this simplification. 
    However, all full numerical calculations we have made show that such a construction minimizes the full search.
    
\begin{figure}
        \centering
        \includegraphics[width=0.9\linewidth]{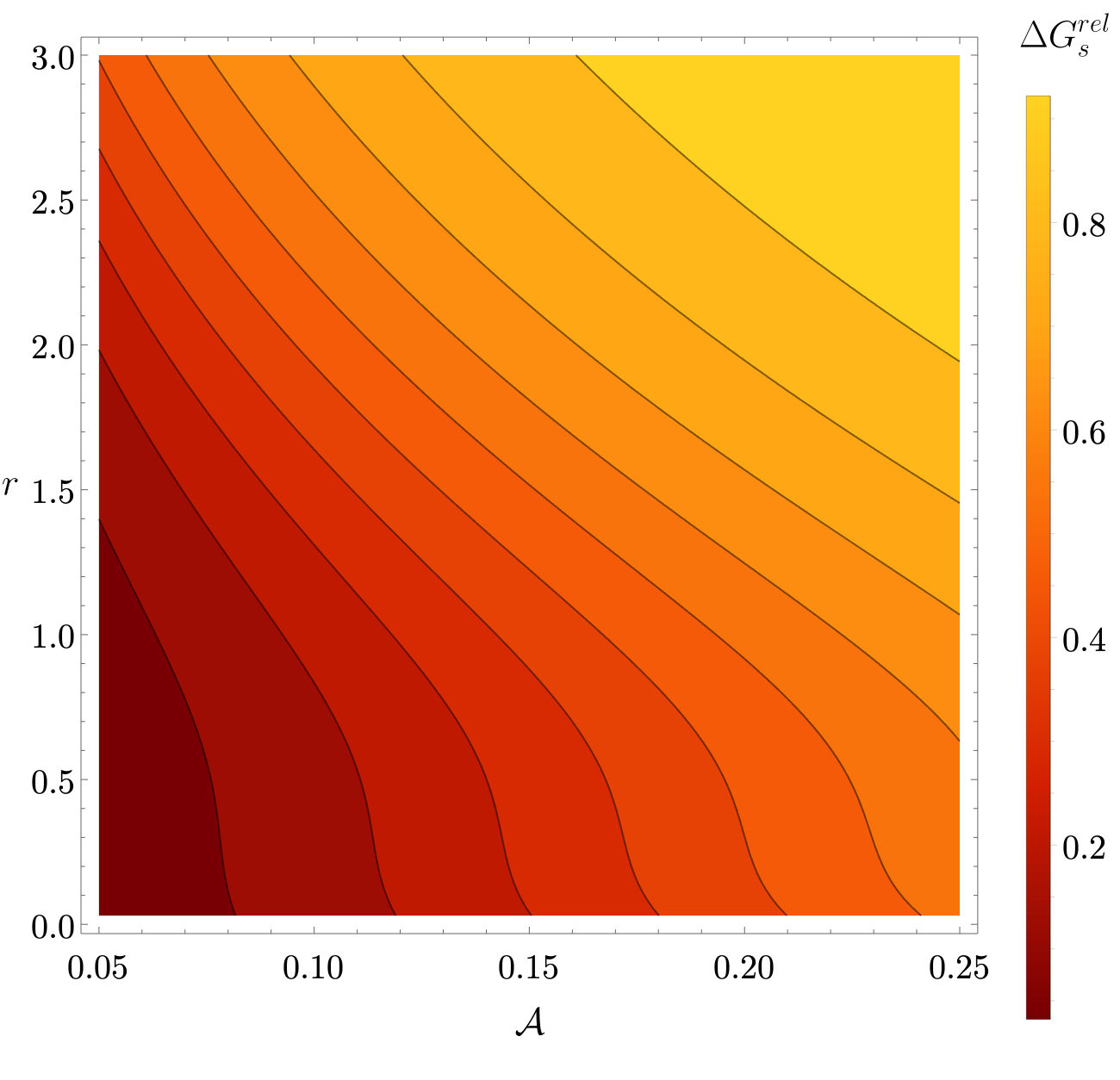}
        \caption{Relative loss of GTE as a function of proper acceleration $\mathcal{A}$ and initial squeezing $r$.}
        \label{fig:relLossGTEsymm}
    \end{figure}

    Solving the equation (\ref{eq:lambda1solve}) for every possible $\alpha$ and initial squeezing of the state $r$ result in the plot in the Fig.~\ref{fig:relLossGTEsymm} which quantifies the relative loss of GTE, i.e., 
    \begin{equation}
        \Delta\text{GTE}^{rel}_s=\frac{G^{res}_\tau(r)-G^{res}_\tau(r_{m})}{G^{res}_\tau(r)}
    \end{equation}
    as function of proper acceleration $\mathcal{A} $.
    The relative loss is the largest for the largest squeezing and increases with the acceleration.
    It reproduces the same qualitative behavior as the purity loss of the multi-mode state under similar geometry~\cite{DD18}.

    We have found a simple procedure for the loss of GTE in the symmetric states. 
    Now, using similar methods, we show the way to quantify the loss in the asymmetric states.

    \subsection{Asymmetric initial state}
As an exemplary asymmetric initial state, we consider bisymmetric states, i.e., states in which two modes can be interchanged without changing the state.
    One can also see them as states squeezed in the third mode differently from the first one.
    A simplified CM of such a family of states can be written in the following form,
        \begin{equation} \label{eq:CMbisym}
            \sigma^p_{bis} = \begin{pmatrix}
                \beta & \zeta &  \zeta' \\
                \zeta & \beta & \zeta'  \\
                \zeta' & \zeta' & \beta'
        \end{pmatrix},
        \end{equation}
    where $\beta$ and $\zeta$ are of the form (\ref{betandzeta}) with squeezing parameter $r_1$.
    The $\beta'$ takes also the form (\ref{betandzeta}), but with different squeezing parameter $r_3$ and $\zeta'$ turns to be
    \begin{equation}
        \zeta'= \begin{pmatrix} \epsilon^+ & 0 \\ 0 & \epsilon^-
        \end{pmatrix}
    \end{equation} with $\epsilon^\pm$ being derived from (\ref{def:epsilon}) with substitution 
    \begin{equation}
        a_i=b(r_1), \quad a_j=b(r_1), \quad a_k=b(r_3). \label{substitution}
    \end{equation}
    
 Transformation of the state from inertial to Rindler frame of reference is achieved with the same quantum channel (\ref{gausschannel}).
    The resulting state is still a Gaussian one, but not pure anymore.
    Thus, one would need to perform a non-trivial optimization problem according to the definition of Gaussian contangle for impure states (\ref{Gmixed}).
    However, we show a solution, analogous to the symmetric case, that requires neither advanced algorithms nor high computing power. 
    
    Similarly, we point out that the family of bisymmetric states has a common eigenbasis with eigenvalues $\gamma_i$ that can be expressed analytically.
    For the sake of clarity of the paper, we have listed them in Appendix \ref{LambaBisSym}. Again, we assume that the symmetric acceleration of observers does not influence symmetry in two symmetric modes. 

\begin{figure}
        \centering
        \includegraphics[width=0.9\linewidth]{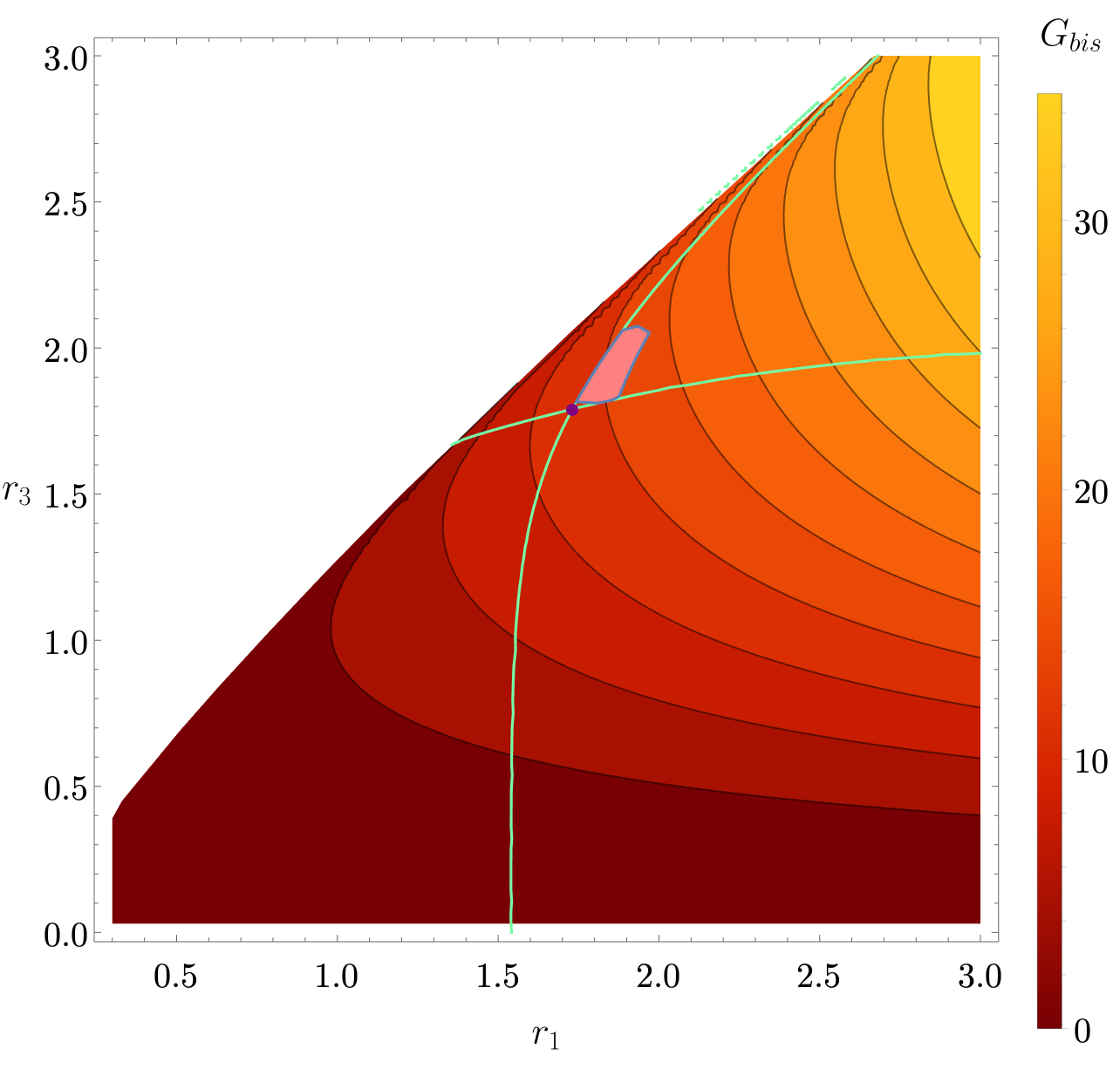}
        \caption{The abscissa axis describes squeezing in the first and second mode $r_1$, and the ordinate axis describes squeezing in the third mode $r_3$. The contour plot shows the value of GTE as a function of $(r_1,r_3)$, the pink region represents the area for which the condition  (\ref{eq:positivity}) with $\mathcal{A}=0.066$ is fulfilled. 
        Green curves are solutions of Eqs. (\ref{bi:lambda1}, \ref{bi:lambda2}).
        For all parameter values we have considered, the crossing of green curves coincides with the bottom left corner of the pink region, yielding the minimum of the Eq. (\ref{Ginf}).}
        \label{fig:GTEbisymmixed}
    \end{figure}

The semi-positivity condition (\ref{semicondition}) considered as a function of initial squeezing $r_1, r_3$ results in "sail"-like area (see pink area in the Fig. \ref{fig:GTEbisymmixed}. 
    The lower-left corner of this allowed area is defined by the intersection of two equations (\ref{eq:lambda1solve}) for two eigenvalues of a bisymmeric state. 
    These equations are 
    \begin{eqnarray}
        \alpha^2(\gamma_1^{(f)}(r_{1},r_{3})-1)-(\gamma_1^{(d)}(r_{1m},r_{3m})-1)&=&0, \label{bi:lambda1} \quad \\
        \alpha^2(\gamma_2^{(f)}(r_{1},r_{3})-1)-(\gamma_2^{(d)}(r_{1m},r_{3m})-1)&=&0,\label{bi:lambda2}
    \end{eqnarray}
    Unfortunately, again we have no analytic proof that these two equations are sufficient for minimization of GTE over impure symmetric states, however, it is the case in all the regimes we consider, which was checked numerically.
    
\begin{figure}
        \centering
        \includegraphics[width=0.9\linewidth]{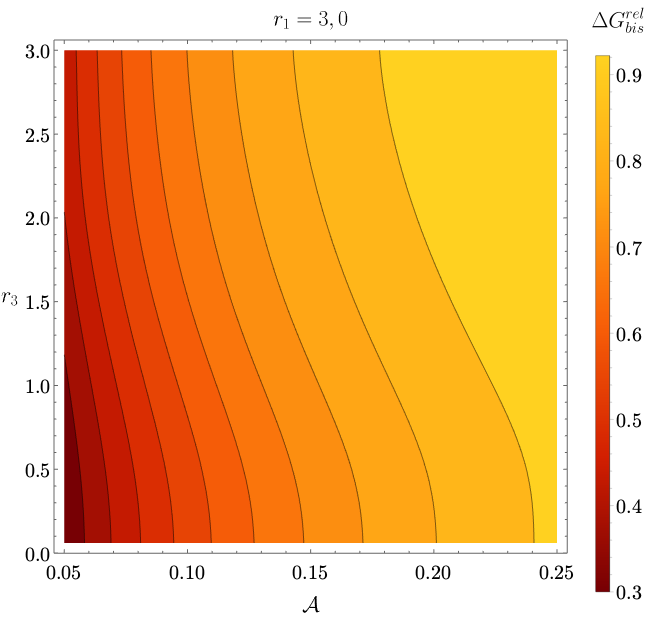}
        \caption{The abscissa axis describes a proper acceleration $\mathcal{A}$ of the observers, and the ordinate axis describes initial squeezing in the third mode $r_3$. All states had initial squeezing in the first and second modes equal to $r_1=3.0$. The contour plot shows the relative loss of GTE, according to (\ref{eq:BisRelLossGTE}).}
        \label{fig:BisRelLossPlot}
    \end{figure}

    Solving the equations (\ref{bi:lambda1}, \ref{bi:lambda2})  with respect to $r_{1m}$ and $r_{3m}$ for every possible $\alpha$ and initial squeezing of a state $r_{1}$, $r_2$ results in the plot in the Fig. \ref{fig:BisRelLossPlot}, which represents the relative loss of GTE, i.e., 
    \begin{equation}
        \Delta\text{GTE}^{rel}_{bis}=\frac{G^{res}_\tau(r_1,r_3)-G^{res}_\tau(r_{1m},r_{3m})}{G^{res}_\tau(r_1, r_3)}. \label{eq:BisRelLossGTE}
    \end{equation}    
    The results are plotted as a function of proper acceleration $\mathcal{A}$ and with $r_1~=~3.0$ chosen to be constant.
    The relative loss is again the largest for the largest squeezing and increases with the acceleration.

   \section{Conclusions} \label{conc}
   We have given a quantitative description of the loss of the genuine tripartite entanglement in CV Gaussian states under influence of acceleration of observers.
   For symmetric states, we have shown that the loss of GTE follows qualitatively the state purity loss as shown in~\cite{DD18}.
   As for the bisymmetric ones, we have found similar behavior, suggesting that acceleration may universally destroy genuine multi-mode entanglement for nonvacuum states, which is further suggested by the previous studies.
   For both of these cases, we have provided a simplified numerical algorithm of calculation of GTE, which greatly simplifies the full numerical minimization procedure developed in previous works.
   As for the outlook, analytic proofs for the simplified numerical procedure can be found and other asymmetric states studied.

	\begin{acknowledgments}
        P. T. G. is financed from the (Polish) National Science Center Grant 2020/36/T/ST2/00065 and supported by the Foundation for Polish Science (FNP).
		Center for Theoretical Physics of the Polish Academy of Sciences is a member of the National Laboratory of Atomic, Molecular and Optical Physics (KL FAMO).
	\end{acknowledgments}
\bibliography{references.bib,library.bib}
\onecolumngrid
    \appendix

    \section{Calculation of GTE} \label{CalculationOfGTE}
     One requires following parameters and functions to quantify genuine tripartite entanglement according to formulas given by Adesso and Illuminati \cite{Adesso06}:
    
     \begin{align}
        s&=\frac{a_2+a_3}{2}, \nn \\
        d&=\frac{a_2-a_3}{2}, \nn \\   
        k_{\pm}&=a^2\pm (s+d)^2, \nn \\
        \delta &= (a-2d-1)(a-2d+1)(a+2d-1)(a+2d+1)(a-2s-1)(a-2s+1)(a+2s-1)(a+2s+1), \nn \\
        D&=2(s-d)-\sqrt{2[2k_{-}^2+2k_+ + |k_-|(k_-^2+8k_+)^{1/2}]/k_+}, \nn \\
        m_{-} &= \frac{ |k_{-}| }{(s-d)^2-1} , \nn \\
        m_+&=\frac{\sqrt{2\left[2a^2(1+2s^2+2d^2)-(4s^2-1)(4d^2-1)-a^4-\sqrt{\delta}\right]}}{4(s-d)} , \nn \\
        Q&=\text{arcsinh}^2\left[\sqrt{m^2(a,s,d)-1}\right]+\text{arcsinh}^2\left[\sqrt{m^2(a,s,-d)-1} \right],
    \end{align}
    
    where $m=m_-$ if $D\leq 0$, and $m=m_+$ otherwise (one has $m_+=m_-$ for $D=0$) and with $Q \equiv G_{\tau}^{1|2} + G_{\tau}^{1|3}$. 
    Here,
    \begin{equation}
        G^{res}_{\tau}(\sigma^p) = \text{arcsinh}^2\big[ \sqrt{a^2_{min} - 1}\big] - Q(a_{min}, s, d), \label{eq:GTE}
    \end{equation}

    \section{Eigenvalues of (bi)symmetric states}
    Here, we provide explicit formulas for symplectic eigenvalues of (bi)symmetric states.
    \subsection{Eigenvalues of symmetric states}\label{LambaSym}
    \begin{align}
        \lambda_1&=\frac{e^{-4 r}+2}{\sqrt{4 \cosh (4 r)+5}}, \quad  \lambda_2=\frac{e^{4 r}+2}{\sqrt{4 \cosh (4 r)+5}}, \nn \\
        \lambda_3&=\frac{2 e^{4 r}+1}{\sqrt{4 \cosh (4 r)+5}}, \quad  \lambda_4=\frac{2 e^{-4 r}+1}{\sqrt{4 \cosh (4 r)+5}}.
    \end{align}
    
    \subsection{Eigenvalues of bisymmetric states}\label{LambaBisSym}
    \begin{eqnarray}
        \gamma_1&=&b_1-\epsilon^+_{12}, \label{l1} \nn \\ 
        \gamma_2&=&\frac{1}{2} \left(\epsilon^+_{12}+b_1+b_3-\sqrt{2 b_1 \epsilon^+_{12}+(\epsilon^+_{12})^2-2b_3 \epsilon^+_{12}+8(\epsilon^+_{13})^2-2b_1 b_3+b_1^2+b_3^2}\right), \label{l2} \nn \\ 
        \gamma_3&=&\frac{1}{2} \left(\epsilon^+_{12}+b_1+b_3+\sqrt{2 b_1 \epsilon^+_{12}+(\epsilon^+_{12})^2-2b_3 \epsilon^+_{12}+8(\epsilon^+_{13})^2-2b_1 b_3+b_1^2+b_3^2}\right), \label{l3} \nn \\ 
        \gamma_4&=&b_1-\epsilon^-_{12}, \label{l4} \nn \\
        \gamma_5&=&\frac{1}{2} \left(\epsilon^-_{12}+b_1+b_3-\sqrt{2 b_1 \epsilon^-_{12}+(\epsilon^-_{12})^2-2b_3 \epsilon^-_{12}+8(\epsilon^-_{13})^2-2b_1 b_3+b_1^2+b_2^2}\right), \label{l5} \nn \\ 
        \gamma_6&=&\frac{1}{2} \left(\epsilon^-_{12}+b_1+b_3+\sqrt{2 b_1 \epsilon^-_{12}+(\epsilon^-_{12})^2-2b_3 \epsilon^-_{12}+8(\epsilon^-_{13})^2-2b_1 b_2+b_1^2+b_3^2}\right), \label{l6}
    \end{eqnarray}\
    where $b_3$ is the local mixedness in the third mode and $\epsilon_{ij}^{\pm}$ are function according to definition (\ref{def:epsilon}) with the same substitution  as in (\ref{substitution}).

    \section{Properties of function of genuine tripartite entanglement}\label{PropsOfGTE}
\begin{figure}[h!]
        \centering
        \includegraphics[width=0.4\linewidth
        ]{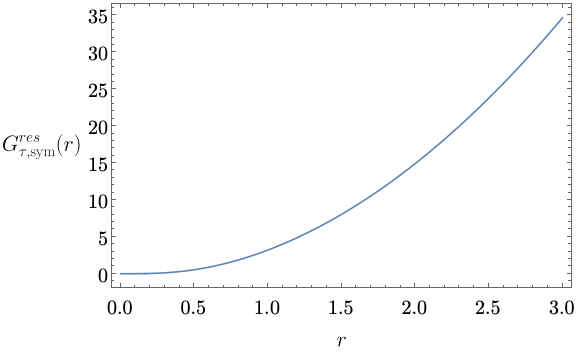}
        \caption{Plot of GTE as a function of symmetric squeezing $r$. }
        \label{fig:GsymPlot}
    \end{figure}

\begin{figure}[h!]
        \centering
        \includegraphics[width=0.4\linewidth]{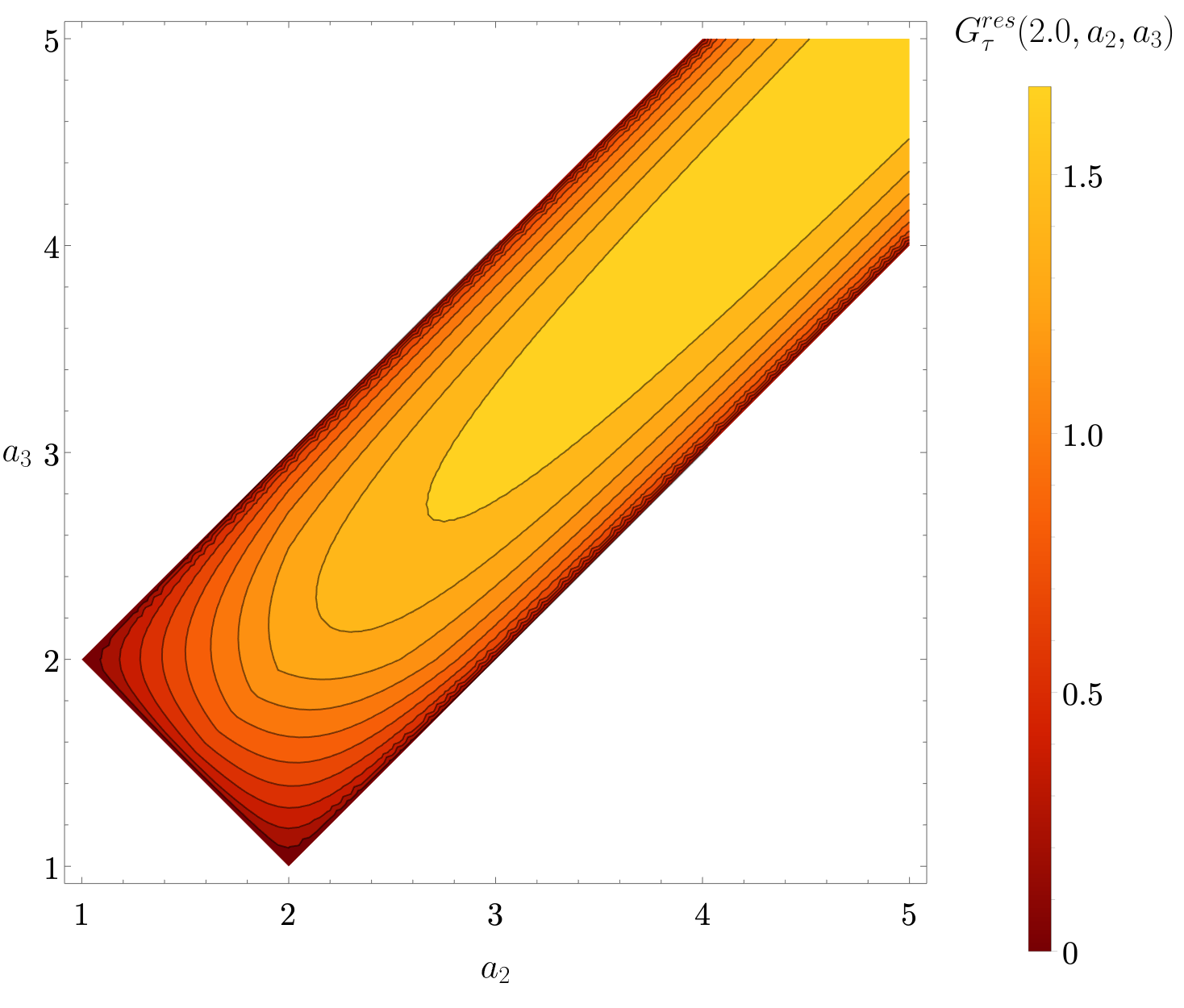}
        \caption{Plot of GTE (\ref{eq:GTE}) as a function of local mixednesses $a_2$ and $a_3$ with constant $a_1=2.0$. The plotted domain is determined by (\ref{eq:triangle}). }
        \label{fig:GTE}
    \end{figure}

    We want to stress the following properties of GTE:
    \begin{itemize}
        \item As squeezing $r$ goes to infinity, the GTE goes to infinity as well. 
        \item Local mixdnesses $a_1, a_2, a_3$ completely determine the value of GTE.
        \item The GTE of 3-mode vacuum states as a function of squeezing $r$ is presented in Fig. \ref{fig:GsymPlot}.
        \item In Fig. \ref{fig:GTE} we present $G^{res}_\tau$ from Eq. (\ref{def:Gres}) as a function of $a_2$ and $a_3$ with fixed $a_1=2.0$.
    \end{itemize}
    
    \section{Two equivalent parameterizations} \label{TwoParametrizations}
    One may prove that $(\ref{eq:CMp})$ and $(\ref{eq:CMvacuum})$ describe the same family of fully symmetric states, using two equivalent parameterizations. If we substitute all the $a_i$ parameters in (\ref{def:epsilon}) with the same $b$ parameter from (\ref{def:b}), we get
    \begin{equation}
        \epsilon_{ij}^{\pm}=\frac{\sqrt{\left(b^2-1\right)^2}\pm\sqrt{9 b^4-10 b^2+1}}{4 \sqrt{b^2}} \label{er23}
    \end{equation}
    We can safely skip all the square roots as we consider $r\in\mathbb{R}_+$ only. By checking that, 
    \begin{equation}
        (b^2-1)=\frac{4}{3}\big(\cosh{(4r)}-1\big),
    \end{equation}
    using hyperbolic identity $\cosh(2x)=\sinh^2(x)+\cosh^2(x)$, we immediately receive 
    \begin{equation}
        (b^2-1)=\frac{8}{9}\sinh^2(2r).
    \end{equation}
    The second expression in the numerator of~\eqref{er23} can be expressed as:
    \begin{equation}
        \sqrt{9b^4 - 10b^2 +1} = \frac{4}{3}\sqrt{\cosh^2(4r)-1}.
    \end{equation}
    Using identity $\cosh^2(x)-\sinh^2(x)=1$ we get
    \begin{equation}
        \sqrt{9b^4 - 10b^2 +1} = \frac{4}{3}\sinh(4r).
    \end{equation}
     After these transformations we receive expressions $z_1$ and $z_2$ from $\epsilon_{ij}^\pm$.

\end{document}